\documentstyle[12pt,epsf]{article}
\newcommand{\postscript}[2] 
{\setlength{\epsfxsize}{#2\hsize}  
\centerline{\epsfbox{#1}}}

\newcommand{\nc}{\newcommand}
\nc{\bg}{B. Grz\c{a}dkowski}
\nc{\lsp}{\;\;\;\;\;\;\;\;}
\nc{\beq}{\begin{equation}}   \nc{\eeq}{\end{equation}}
\nc{\bea}{\begin{eqnarray}}   \nc{\eea}{\end{eqnarray}}
\nc{\baa}{\begin{array}}      \nc{\eaa}{\end{array}}
\nc{\bit}{\begin{itemize}}    \nc{\eit}{\end{itemize}}
\nc{\ben}{\begin{enumerate}}  \nc{\een}{\end{enumerate}}
\nc{\bce}{\begin{center}}     \nc{\ece}{\end{center}}
\nc{\ra}{\rightarrow}
\nc{\non}{\nonumber}
\nc{\barx}{\bar{x}}
\nc{\pbarn}{\rm pb}
\nc{\fmbarn}{\rm fb}
\nc{\re}{\hbox {Re}}
\nc{\mev}{\hbox {MeV}} \nc{\gev}{\;\hbox {GeV}} \nc{\tev}{\;\hbox {TeV}}
\nc{\etal}{{\it et al.}}
\def\gesim{\lower0.5ex\hbox{$\:\buildrel >\over\sim\:$}} 
\def\lesim{\lower0.5ex\hbox{$\:\buildrel <\over\sim\:$}} 
\nc{\prd}[3]{{\it Phys.\ Rev.}\ {{\bf D{#1}} (#2), #3}}
\nc{\prl}[3]{{\it Phys.\ Rev.\ Lett.}\ {{\bf {#1}} (#2), #3}}
\nc{\plb}[3]{{\it Phys.\ Lett.}\ {{\bf B{#1}} (#2), #3}}
\nc{\npb}[3]{{\it Nucl.\ Phys.}\ {{\bf B{#1}} (#2), #3}}
\nc{\ptp}[3]{{\it Prog.\ Theor.\ Phys.}\ {{\bf {#1}} (#2), #3}}
\nc{\zfp}[3]{{\it Z.\ Phys.}\ {{\bf C{#1}} (#2), #3}}
\nc{\mpla}[3]{{\it Mod.\ Phys.\ Lett.}\ {{\bf A{#1}} (#2), #3}}
\nc{\rmp}[3]{{\it Rev.\ Mod.\ Phys.}\ {{\bf {#1}} (#2), #3}}
\nc{\ijmpa}[3]{{\it Int.\ J.\ of\ Mod.\ Phys.}\
               {{\bf A{#1}} (#2), #3}}
\nc{\ttbar}{t\bar{t}}         \nc{\bbbar}{b\bar{b}}
\nc{\twbdec}{t\to W^+ b}
\nc{\tbwbdec}{\bar{t}\to W^- \bar{b}}
\nc{\epem}{e^+e^-}            \nc{\eett}{\epem \to \ttbar}
\nc{\sigeett}{\sigma_{e\bar{e}\to\ttbar}}  
\nc{\tbar}{\bar{t}}           \nc{\bbar}{\bar{b}}
\nc{\mt}{m_t}                 \nc{\mts}{m_t^2}
\nc{\lp}{\ell^+}              \nc{\lm}{\ell^-}
\nc{\epsl}{\epsilon_L}        \nc{\leff}{L\epsilon_{\ell\ell}}
\nc{\cp}{$C\!P$}
\nc{\sig}{\frac{1}{\sigma}\frac{d\sigma}{d\phi}}
\nc{\et}{E_T}
\nc{\proc}{p \bar{p} \ra \ttbar +X \ra l^+ l^- + X \;/\; l^\pm + X}

\begin{document}
\pagestyle{empty} \setlength{\footskip}{2cm}
\setlength{\oddsidemargin}{0.5cm} \setlength{\evensidemargin}{0.5cm}
\renewcommand{\thepage}{-- \arabic{page} --}
\def\mib#1{\mbox{\boldmath $#1$}}
\def\bra#1{\langle #1 |}      \def\ket#1{|#1\rangle}
\def\vev#1{\langle #1\rangle} \def\dps{\displaystyle}
% -------------------------------------------------------------------
   \def\thebibliography#1{\centerline{REFERENCES}
   \list{[\arabic{enumi}]}{\settowidth\labelwidth{[#1]}\leftmargin
   \labelwidth\advance\leftmargin\labelsep\usecounter{enumi}}
   \def\newblock{\hskip .11em plus .33em minus -.07em}\sloppy
   \clubpenalty4000\widowpenalty4000\sfcode`\.=1000\relax}\let
   \endthebibliography=\endlist
   \def\sec#1{\addtocounter{section}{1}
       \section*{\normalsize\bf\arabic{section}. #1}\vspace*{-0.3cm}}
% -------------------------------------------------------------------
\vspace*{-1cm}\noindent
\hspace*{10.cm}UW-IFT-05/97\\
\hspace*{10.cm}MPI-PhT/97-33\\
%\hspace*{10.cm}(hep-ph/9706489)\\
%\hspace*{10.cm}version: \today \\

\vspace*{.5cm}

\begin{center}
{\large\bf $\mib{C}\!\mib{P}$ Violation, Top Quarks and the Tevatron Upgrade}
\end{center}

\vspace*{1.25cm}
\begin{center}
\renewcommand{\thefootnote}{*)}
{\sc B.
GRZ\c{A}DKOWSKI$^{\: a),\: }$}\footnote{E-mail address:
\tt bohdan.grzadkowski@fuw.edu.pl}
\renewcommand{\thefootnote}{**)}
{\sc B. LAMPE$^{\: b),\: }$}\footnote{E-mail address:
\tt bol@mppmu.mpg.de}
\renewcommand{\thefootnote}{***)}
{\sc K.J. ABRAHAM$^{\: c),\: }$}\footnote{E-mail address: 
\tt abraham@unpsun1.cc.unp.ac.za}

\vspace*{1.2cm}
\centerline{\sl $a)$ Instytut Fizyki Teoretycznej,\ Uniwersytet Warszawski}
\centerline{\sl Ho\.za 69, PL-00-681 Warszawa, POLAND} 

\vskip 0.3cm
\centerline{\sl $b)$ Max-Planck-Institute f\"ur Physik und Astrophysik}
\centerline{\sl F\"ohringer Ring 6,\ D-80805 M\"unchen, GERMANY}

\vskip 0.3cm
\centerline{\sl $c)$  Department of Physics, University of Natal}
\centerline{\sl Pietermaritzburg, SOUTH AFRICA}

\end{center}

\vspace*{1.5cm}
\centerline{ABSTRACT}

\vspace*{0.4cm}
\baselineskip=20pt plus 0.1pt minus 0.1pt
In order to observe a signal of possible $C\!P$ violation in top-quark
couplings,
we study top-quark production and decay under the conditions
of the Tevatron upgrade.
Transverse energy asymmetries sensitive to $C\!P$ violation are
defined.
Applying the recently proposed optimal method,
we calculate  the
statistical significance for the direct observation of $C\!P$ violation
in the
production and subsequent decay of top quarks.

\vfill
\newpage
%--------------------------------------------------------------------
\renewcommand{\thefootnote}{\sharp\arabic{footnote}}
%--------------------------------------------------------------------
\pagestyle{plain} \setcounter{footnote}{0}
\baselineskip=21.0pt plus 0.2pt minus 0.1pt

\sec{Introduction}

The top quark, thanks to its huge mass, is expected to provide a
good opportunity to study physics beyond the Standard Model (SM). Indeed,
as many authors have pointed out $[1\ -\ 8]$,  \cp\ violation in the 
combined process of top-quark production
and decay could be a useful signal for possible non-standard interactions.
This is because ({\it i}) the \cp\ violation in the top-quark
couplings induced within the SM
is negligible and ({\it ii}) a lot of
information on the top quark is transferred to the secondary leptons
without getting obscured by the hadronization effects.

In this letter we will consider two types of semi-inclusive 
processes: first, the process in which both 
top-quarks decay semileptonically,   
\begin{equation}
p \bar p \rightarrow t \bar t \rightarrow l^+ l^- X
\, ,
\label{pr1}
\end{equation}
and secondly those processes in which only one of them 
decays semileptonically, 
\begin{equation}
p \bar p \rightarrow t \bar t \rightarrow l^+ X 
\qquad 
p \bar p \rightarrow t \bar t \rightarrow l^- X         \, .              
\label{pr2}  
\end{equation}
The latter processes (\ref{pr2}) are particularly interesting because they 
have a better statistics and give the best signature for the top-quark 
identification. 

The main production mechanism for the top-quark production 
at the Tevatron is $q\bar q$--annihilation to top quarks where 
the quark and the anti-quark stem from a high energy proton and 
anti-proton, respectively. In principle there are also gluon 
processes to consider but at Tevatron energies 
they give only a small contribution to the cross section of the 
order of 10\%, and less than 1\% to the $C\!P$ sensitive observables 
considered here. Therefore they will be neglected in the calculations. 

We will apply the usual CDF cuts in our analysis \cite{topdisc}. 
For example, 
a $p_T$--cut of 5 GeV for all leptons and a rapidity cut of 
3 for all particles will be introduced. 
It was checked that these cuts  
do not induce fake effects in the \cp-sensitive observables. 
We will adopt the Tevatron upgrade energy $\sqrt{s}=2.0\tev$.
Two options for the luminosity will be considered here; so called
``TeV-33'' defined as $L=30\; \fmbarn^{-1}$ and the Tevatron Run II with
$L=2\; \fmbarn^{-1}$ at the same energy.

%%%%%%%%%%%%%%%%%%%%%%%%%%%%%%%%%%%%%%%%%%%%%%%%%%%%%%%%%%%%%%%%%%%%%%%%%%%%%%

\sec{The optimal method}

In order to be as sensitive as possible 
to \cp-violating couplings we will adopt here
the recently proposed optimal procedure~\cite{opt-96} for the data analysis.
Let us briefly summarize the main points of this method.
Suppose we have a cross section
\begin{equation}
\sig=\sum_i c_i f_i(\phi)
\label{sta}
\end{equation}
where the $f_i(\phi)$ are known functions of the location in
final-state phase space $\phi$ and the $c_i$ are model-dependent
coefficients. $\sigma$ is the integrated cross section 
$\sigma =\int \frac{d\sigma}{d\phi} d\phi$. 
In this paper we will restrict ourselves to a case for which
all of the $c_i$ 
are small except for $c_1=1$ ($f_1(\phi)$ will be the
SM contribution, the other $c_i$ will parameterize beyond the SM physics).
The ultimate goal would be to determine  the $c_i$'s. It can be done
by using appropriate weighting functions $w_i(\phi)$ such that $\int
w_i(\phi)\sig(\phi)d\phi=c_i$. Generally, different choices for
$w_i(\phi)$ are possible, but there is a unique choice such that the
resultant statistical error is minimized. Such functions are given by
\begin{equation}
w_i(\phi)=\sum_j \frac{X_{ij}f_j(\phi)}{{\sig}(\phi)}\,,
\label{X_def}
\end{equation}
where $X_{ij}$ is the inverse matrix of $M_{ij}$ which is defined as
\begin{equation}
M_{ij}\equiv \int {f_i(\phi)f_j(\phi)\over{\sig}(\phi)}
d\phi\,.
\label{M_def}
\end{equation}
When we take these weighting functions, the statistical uncertainty
of $c_i$ becomes
\begin{equation}
{\Delta}c_i=\sqrt{V_{ii}},
\end{equation}
for the covariance matrix $V$ defined as
\beq
V_{ij}=\frac{1}{N}\int w_i(\phi)w_j(\phi)\sig\;d\phi,
\label{cov_mat}
\eeq
where $\sigma$ and $N$ stand for the total cross section and the total
number of observed events, respectively.

Since we assume that non-standard interactions do not alter the SM pattern
radically, we will keep only linear terms in the $c_i$ (those which
arise from an interference with SM contributions). 
Within this approximation, 
the expression $\sig$ in Eqs. (\ref{X_def}-\ref{cov_mat}) can be  
replaced by the SM contribution $f_1$. 

%%%%%%%%%%%%%%%%%%%%%%%%%%%%%%%%%%%%%%%%%%%%%%%%%%%%%%%%%%%%%%%%%%%%%%%%%%%%%%

\sec{\cp-violating transverse energy asymmetries}

We will assume that all non-standard effects in the production
process $q\bar q \rightarrow t\bar t$ 
can be represented by the gluon exchange in
the $s$-channel with the following effective coupling:
\footnote{Two other possible
form factors do not contribute in the limit of zero parton mass.}
\begin{equation}
{\Gamma}^\mu (g^{\ast} \rightarrow t\bar t)=g_{s}\bar{u}(p_t)
\biggl[\,\gamma^\mu (F_1^L P_L+F_1^R P_R)
-\frac{i \sigma^{\mu \nu}(p_t+p_{\bar{t}})_\mu}{\mt}(F_2^L P_L+F_2^R P_R)\,\biggr]v(p_t),
\label{gtt}
\end{equation}
where $g_{s}$ is the strong coupling constant, $P_{L/R}\equiv(1\mp
\gamma_5)/2$ and colour degrees of freedom have been omitted. 
The SM vertex is given by $F_1^R=F_1^L=1$ and  $F_2^R=F_2^L=0$. 
A non-zero value of 
$F_2^L-F_2^R$ is a signal of \cp\ violation.

For the on-shell $W$, 
we will adopt the following parameterization of the $tbW$ vertex
suitable for the decays $\twbdec$ and $\tbwbdec$:
\beq
{\Gamma}^{\mu}(\twbdec )=-{g\over\sqrt{2}}V_{tb}\:
\bar{u}(p_b)\biggl[\,\gamma^{\mu}(f_1^L P_L +f_1^R P_R)
-{{i\sigma^{\mu\nu}k_{\nu}}\over M_W}
(f_2^L P_L +f_2^R P_R)\,\biggr]u(p_t),
\eeq
\beq
\bar{\Gamma}^{\mu}(\tbwbdec )=-{g\over\sqrt{2}}V_{tb}^*\:
\bar{v}(p_t)\biggl[\,\gamma^{\mu}(\bar{f}_1^L P_L +\bar{f}_1^R P_R)
-{{i\sigma^{\mu\nu}k_{\nu}}\over M_W}
(\bar{f}_2^L P_L +\bar{f}_2^R P_R)\,\biggr]v(p_b),
\eeq
where $g$ is the SU(2) gauge-coupling constant, $V_{tb}$ is the $(tb)$ element of the Kobayashi-Maskawa
matrix and $k$ is momentum of the $W$. For the SM tree-level interactions has
$f_1^L=\bar{f}_1^L=1$ and
$f_1^R=f_2^L=f_2^R=\bar{f}_1^R=\bar{f}_2^L=\bar{f}_2^R=0$. 
Again,
because $W$ is on shell, there are two additional form factors 
which do not
contribute. One can show that \cite{cprelation} 
\begin{equation}
f_1^{L,R}=\pm\bar{f}_1^{L,R},\lsp f_2^{L,R}=\pm\bar{f}_2^{R,L},
\end{equation}
where upper (lower) signs are those for $C\!P$-conserving
(-violating) contributions. Therefore any observable 
sensitive to \cp-violation in the top-quark decay 
must be proportional to $f_1^{L,R}-
\bar{f}_1^{L,R}$ or $f_2^{L,R}-\bar{f}_2^{R,L}$\footnote{For $m_b=0$
 only $\bar{f}_2^L$ and $f_2^R$ interfere with the SM, therefore
in the leading order in the nonstandard couplings, 
only terms proportional to $\bar{f}_2^{L}-f_2^{R}$ will appear in the cross
section.}.

The matrix element for the combined production and decay of 
top quarks via $q\bar q$--annihilation, $q\bar q \ra t \bar t \ra l^+l^- +
\dots$ has been obtained by the algebraic computer program FORM~\cite{form}.
For the massless $b$ quark the resulting expression  consists of a 
\cp-conserving SM piece plus two terms linear in the \cp-violating couplings 
$c_P$ and $c_D$ corresponding to \cp\ violation in the production and in
the decay process, respectively: 
\beq
c_{SM}=1 \lsp c_P=\frac{1}{2}\re(F_2^L-F_2^R) \lsp
c_D=\frac{1}{2}\re(f_2^R - \bar f_2^L) \, .
\eeq
The notation of Sect. 2 has been used here with the indices 
$1={\hbox {SM}}$, $2={\hbox {P}}$(roduction) and $3={\hbox {D}}$(ecay). 
Hereafter we assume that $m_b=0$ and 
all non-standard interactions violate \cp,
i.e. $F_2^L=-F_2^R$ and $f_2^R =- \bar f_2^L$.
The matrix element squared will be then 
convoluted with the Morfin and Tung \cite{morfin}
parton distributions (the 'leading order' set from the 'fit sl').
Numerical results will be 
obtained using the Monte Carlo package RAMBO \cite{kleiss}.

In general, both \cp\ violating couplings $c_P$ and $c_D$ may be 
present. Therefore 
let us discuss how to observe a combined signal of $C\!P$ violation
emerging from these couplings. 
There are many, more or less efficient observables to be 
considered in the discussed process. In this letter 
only those related to the transverse energy of 
the muon, will be considered.
A possible \cp-sensitive asymmetry 
for the process (\ref{pr1}) is for instance the 
transverse-energy ($E_T^{\pm}$) asymmetry for the muons:
\begin{equation} 
A_T^{\mu}=
{\sigma (E_T^- > E_T^+)-\sigma (E_T^+ > E_T^-) 
\over 
\sigma (E_T^- > E_T^+)+\sigma (E_T^+ > E_T^-)}.
\label{lll}
\end{equation} 
A non-zero observation of $A_T^{\mu}$ would prove \cp\ violation in
the production or in the decay of the top quarks. 

The asymmetry has been discussed in Ref.~\cite{SP,bernr}. 
It is revisited here because we present a detailed numerical analysis
applicable for the 
Tevatron upgrade conditions. 
Furthermore we are able to give results both for 
\cp\ violation from the decay and from the production vertex whereas in 
Ref. \cite{SP} only \cp\ violation from the production side was 
considered.  
The result for the expected \cp-violating effect is shown in Fig. 1  
as a function  of the couplings $c_P$ and $c_D$. 
It is seen that $c_P$ gives larger effects by about a 
factor of three. The effects may be remarkably large, 
however since we have neglected terms quadratic in $c_i$,
one should not consider $c_i$'s greater than $0.2$ in order to retain
precision at the level of several per cent.
In general, \cp\ violation may be
present both in the 
production and decay of top quarks. Therefore one should take 
an appropriately weighted sum of the two curves in Fig. 1.  
Unfortunately, lacking a real theory of \cp\ violation the weights 
are not known. 

\begin{figure}
\postscript{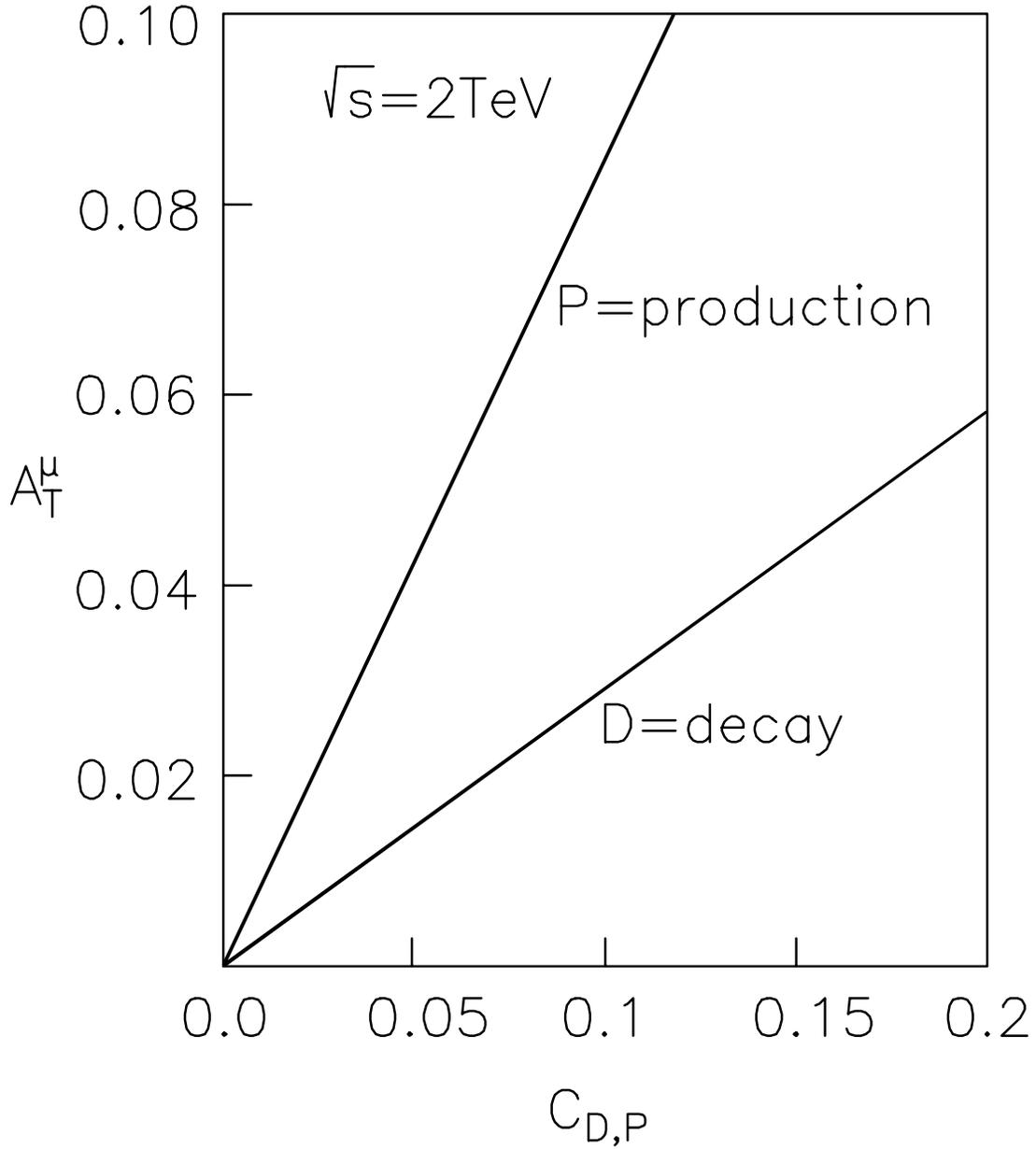}{1.0}
\bigskip
\caption{Transverse energy asymmetry as defined 
in Eq. (\protect\ref{lll}) 
for muons from semileptonic 
top-quark decays, as induced by \cp\ violating couplings 
$c_P$ and $c_D$ 
at the production and decay vertex, respectively.}
\end{figure}

It should be noted that 
among the transverse energy asymmetries 
the muon $E_T$--asymmetry is the most efficient to tag for 
\cp-violating effects, not only because muons have much clearer 
signatures than other top-quark decay products but also 
due to the structure of the \cp-violating matrix elements. 
For example, if one works
out the matrix element for $c_P=c_D=0.1$ one obtains a $A_T^{\mu}=+11.3$\% 
(as seen in Fig. 1) whereas $A_T^{W}$, $A_T^{\nu}$, $A_T^{b}$, etc. 
are all smaller, $A_T^{W}=5.3$\%, $A_T^{\nu}=-4.8$\% and $A_T^{b}=-5.2$\%. 
This feature is independent of the values chosen for $c_P$ and $c_D$. 

It is worth noting that the transverse-energy asymmetry $A^\mu_T$ 
is identical to the normalized expectation value 
$<w_E>=\int w_E \sig d\phi$ of the following weighting function:  
\beq
w_E(\et^+,\et^-)\equiv\frac{\et^- - \et^+}{|\et^- - \et^+|}.
\eeq
$A^\mu_T$ can be decomposed according 
to 
\beq 
A^\mu_T=<w_E>=\int w_E \sig d\phi=
c_P \int w_E f_P d\phi +c_D \int w_E f_D d\phi 
\equiv c_P A_P +c_D A_D 
\label{deee} 
\eeq 
with $A_P$ and $A_D$ calculated to be $A_P=0.845$ and $A_D=0.285$. 
Since $A_P>A_D$ we can conclude that it is easier to observe \cp\ violation
in the production process.~\footnote{The same phenomenon has been noticed
for the process $\epem \ra \ttbar \ra l^+\;l^-\;X$ and 
$\epem \ra \ttbar \ra l^\pm\;X$, see Ref.~\cite{BGZH}. It might have been
anticipated as a consequence of the fact that the components of
$ (p_t + p_{\bar{t}})/ m_t $
(production) are usually greater than those of 
$k/m_w$ 
(decay).}
For the number of dilepton events denoted by $N_{ll}$, the statistical 
significance for $A^\mu_T$ determination is given by
\beq
N_{SD}^T \equiv |A^\mu_T|\sqrt N_{ll}=|c_P A_P +c_D A_D|\sqrt N_{ll}
\eeq
The expected number of events in the dilepton mode~\cite{upgrade}, 
$80$ and $1200$ at 
the integrated luminosity $L=2 \; \hbox{and} \; 30 \; \fmbarn^{-1}$, respectively,
allows an observation of  
the $3\sigma$ effect providing the following relations are satisfied:
\bea
|2.5 c_P + 0.9 c_D|\geq 1 & {\rm for} & L=2\; \fmbarn^{-1}\\
|9.8 c_P + 3.3 c_D|\geq 1 & {\rm for} & L=30\; \fmbarn^{-1}
\eea
So, we can observe that even $L=2\; \fmbarn^{-1}$ allows for an observation of
$c_P=c_D=.3$ at $\sqrt{s}=2 \tev$.

Since the dilepton events are  
relatively rare and difficult to 
identify we shall discuss
another observable here:
\begin{equation}
A_{cut}^{\mu}(E_{T cut})=
{\sigma^-(E_T^- > E_{T cut})-\sigma^+(E_T^+ > E_{T cut})
\over 
\sigma }
\label{112}
\end{equation} 
which can be used for the processes Eq. (\ref{pr2}). 
The $\sigma$ in (\ref{112}) denotes the 
integrated cross section with no 
cuts except for the standard CDF cuts,
Note that the transverse-energy-spectrum asymmetry 
$1/\sigma(d\sigma^+/d\et-d\sigma^-/d\et)$ 
may be obtained from $A_{cut}^{\mu}(E_{T cut})$ just by 
differentiation with respect to $E_{T cut}$. 
The dependence of $A_{cut}^{\mu}$ as a function of $E_{T cut}$ 
is shown in Fig. 2 for two sets of the couplings, ($c_P=0,\;\;c_D=0.1$) 
and ($c_P=0.1,\;\;c_D=0$). 
>From Fig. 2 one can read off 
the $E_{T cut}$--region where the 
transverse-energy-spectrum asymmetry is maximal:
$\et=50\gev$ and $\et=35\gev$ for \cp\ violation
in the production and decay, respectively.

Again, it is seen that effects of \cp-violation in the production process
are more pronounced.

As it is seen from our analysis it may happen that the values of $c_P$ and
$c_D$ would conspire in such a way that the asymmetries discussed would be
very small; \cp\ violation in the production and decay would cancel each
other. Therefore it would be very useful to be able to disentangle \cp\
violation in the production and decay. The method of optimal observables
introduced in the Chapter 2 provides the desired strategy. In the next
chapter we present our numerical results for the separate determination of
\cp\ violation in the production and decay.

\begin{figure}
\postscript{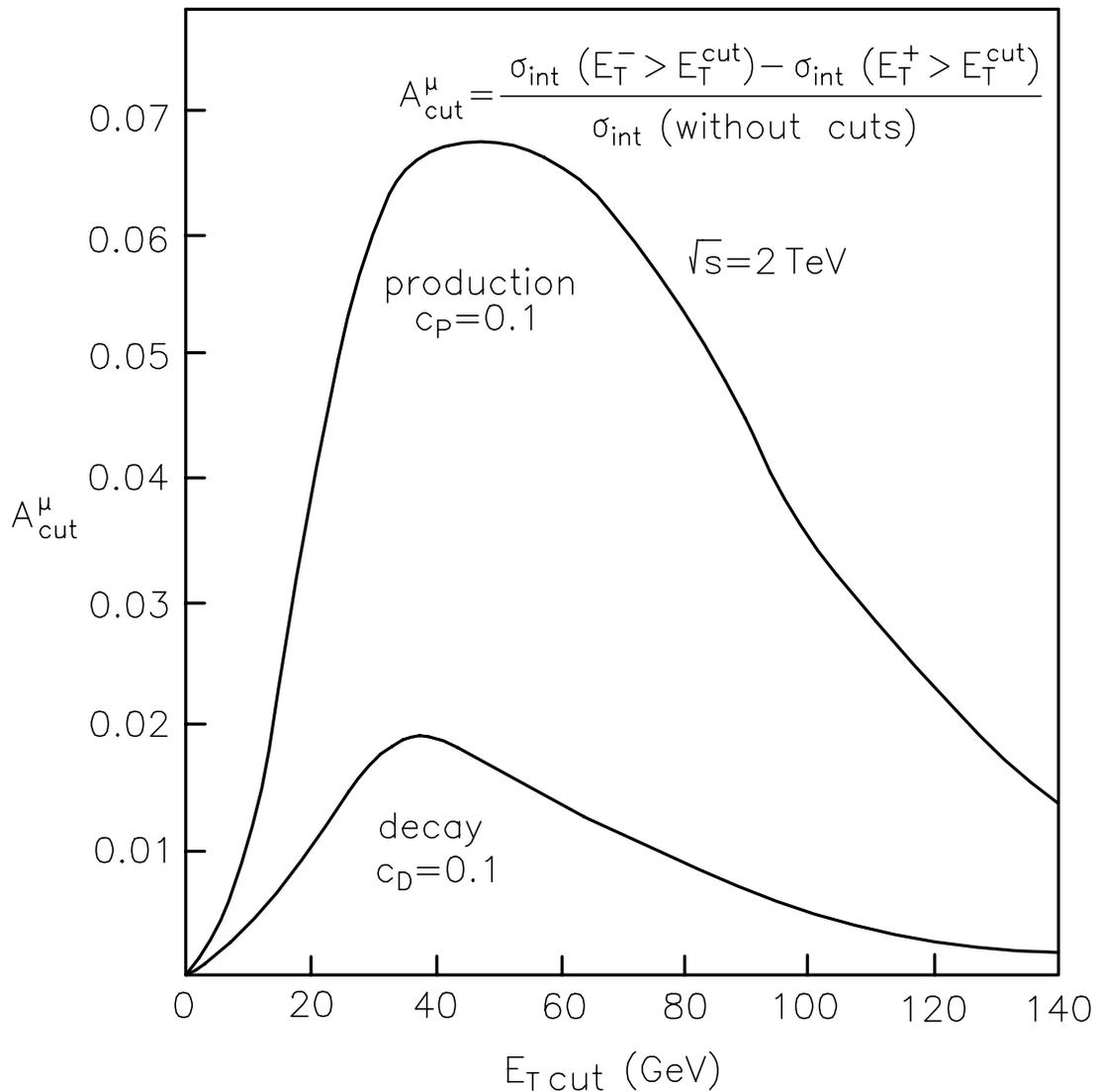}{1.0}
\bigskip
\caption{
Transverse energy asymmetry as defined in Eq. (\protect\ref{112})  
for muons from semileptonic  
top-quark decays, as induced by \cp\ violating couplings $c_P$  
and $c_D$ at the production and decay vertex, respectively.}
\end{figure}

%%%%%%%%%%%%%%%%%%%%%%%%%%%%%%%%%%%%%%%%%%%%%%%%%%%%%%%%%%%%%%%%%%%%%%%%%%%%%%

\sec{The optimal observables} 

It may be interesting to compare the statistical significance for the 
transverse-energy asymmetry with the one calculated for the optimal 
(for a detection of non-zero $c_{P}$ or $c_{D}$) observables  
defined in Sec.2. 
We shall consider the transverse-lepton-energy spectrum in the final state defined 
through the detection of high-$\et l^+$ + jets
with appropriate cuts
included. The spectrum is sensitive to $c_P$ and
$\re(f_2^R)$:
\beq
\frac{1}{\sigma} \frac{d \sigma}{dE_T^+}=f_1^+(E_T^+)+c_P f_P^+(E_T^+)
+\re(f_2^R) f_D^+(E_T^+),
\eeq
where $\sigma$ denotes the cross section for the process $p \bar{p} \ra 
l^+ \; \hbox{jets}$ and $f^+$'s are known functions of $E_T^+$. We 
have assumed that all the non-standard interactions violate \cp. 

We have obtained the 
following relevant entries for the matrix 
$M$~\footnote{
Since the discussed form factors enter appropriate vertices multiplied by
the $\ttbar$ or $W^+$ momentum, $M_{ij}$ depend 
on the proton energy. For example, with protons of energy 
$1.5\tev$ one gets $M_{DP}=-0.125$, $M_{PP}=0.33$ and $M_{DD}=0.050$. 
It could be verified that the precision of $c_P$ and $f_2^R$ determination
increases with the proton energy.}:
$$
M_{DP}=0.17 \lsp M_{PP}=0.72 \lsp M_{DD}=0.048
$$
Now, the optimal weighting 
functions can be obtained. The statistical errors $\Delta c_i$ for the 
determination of $c_{P}$ and  $\re(f_2^R)$ are the following:
\beq
\Delta c_P = \sqrt{\frac{M_{DD}}{N_l \Delta}}=\frac{3.35}{\sqrt{N_l}} \lsp 
\Delta \re(f_2^R) = \sqrt{\frac{M_{PP}}{N_l \Delta}}=\frac{13.04.}{\sqrt{N_l}} ,
\eeq
where $\Delta\equiv M_{DD}M_{PP}-M_{DP}^2$ and $N_l$ stands for the total number of 
single lepton events. An analogous procedure leads to $\Delta c_P$ and 
$\Delta \re (\bar{f}_2^L)$ from $l^-$ energy spectrum. Since both
distributions are statistically independent, we can combine them to receive
$\Delta c_P$ and $\Delta c_D$
\beq
\Delta c_P =\frac{2.37}{\sqrt{N_l}} \lsp 
\Delta c_D =\frac{18.43}{\sqrt{N_l}}
\label{errors}
\eeq
In order to estimate the power of the optimal observables
we need to calculate the statistical significance,
$N_{SD}^{P,D}\equiv |c_{P,D}|/\Delta c_{P,D}$  for their experimental
determination:
\beq
N_{SD}^{P}=\frac{|c_P|}{2.37}\sqrt{N_l} \lsp
N_{SD}^{D}=\frac{|c_D|}{18.43}\sqrt{N_l}.
\eeq
The expected number of single-leptonic events (1 b-quark
tagged)~\cite{upgrade} is
$1300$ and $20,000$ for $L=2 \; {\hbox {and}}\; 30\;\fmbarn^{-1}$, respectively.
In Table 1 we show the minimal values for $c_P$ and $c_D$ necessary to
observe $3\sigma$ effects.
\begin{table}
\vspace*{-0.4cm}
\bce
\begin{tabular}{||c|c|c||}
\hline\hline
$L[\fmbarn^{-1}]$&2&30\\
\hline
$|c_P|$&0.20&0.05\\
\hline
$|c_D|$&1.50&0.40\\
\hline\hline
\end{tabular}\\
\vspace*{0.3cm}
%{\bf -- Table 1 --}
\ece
\vspace*{-0.5cm}
\caption{The minimal values for $c_P$ and $c_D$ necessary to observe \cp\
violation in the single-lepton mode at the $3\sigma$ level for $L=2, \; 30
\; \fmbarn^{-1}$.} 
\end{table}

As it has already been noticed it will be much easier to observe \cp\ violation in the
production process; even at $L=2\; \fmbarn^{-1}$, $c_P=0.2$ will be seen at
the $3\sigma$ level.

The transverse-double-lepton-energy spectrum allows for
independent~\footnote{To obtain statistically independent
determination of $c_{P,D}$  one needs to 
discard double leptonic events while measuring the single-lepton-energy
spectrum.}  $c_{P,D}$ determination:
\beq
\frac{1}{\sigma} \frac{d^2 \sigma}{dE_T^+\;dE_T^-}=
f_1^\pm(E_T^+,E_T^-)+c_P f_P^\pm(E_T^+,E_T^-)
+c_D f_D^\pm(E_T^+,E_T^-),
\eeq
where $\sigma$ stands for the cross section for the process 
$p \bar{p} \ra l^+ \; l^- \; \hbox{jets}$ and $f^\pm$'s are known functions
of $E_T^+$ and $E_T^-$. In this case
$M_{ij}$ are the following:
\beq
M_{DP}=\pm 0.57 \lsp M_{PP}=3.19 \lsp M_{DD}=0.13.
\eeq
The statistical significance for $c_{P,D}$ determination read:
\beq
N_{SD}^P=\frac{|c_P|}{1.17}\sqrt{N_{ll}} \lsp
N_{SD}^D=\frac{|c_D|}{5.76}\sqrt{N_{ll}} 
\eeq
Adopting the anticipated number of dileptonic events~\cite{upgrade},
$N_{ll}=80 \; \hbox{and} \; 1200$ for the luminosity $L=2 \; \hbox{and} \; 30\;
\fmbarn^{-1}$ we present in Table 2 the minimal values for $c_P$ and $c_D$
necessary to observe $3\sigma$ effects.
\begin{table}
\vspace*{-0.4cm}
\bce
\begin{tabular}{||c|c|c||}
\hline\hline
$L[\fmbarn^{-1}]$&2&30\\
\hline
$|c_P|$&0.39&0.10\\
\hline
$|c_D|$&1.93&0.50\\
\hline\hline
\end{tabular}\\
\vspace*{0.3cm}
%{\bf -- Table 2 --}
\ece
\vspace*{-0.5cm}
\caption{The minimal values for $c_P$ and $c_D$ necessary to observe \cp\
violation in the dilepton mode at the $3\sigma$ level for $L=2, \; 30 \;
\fmbarn^{-1}$.} 
\end{table}
It is seen from the table that single-leptonic modes are more
promising as signals 
of non-standard and \cp-violating physics than dilepton ones.

%%%%%%%%%%%%%%%%%%%%%%%%%%%%%%%%%%%%%%%%%%%%%%%%%%%%%%%%%%%%%%%%%%%%%%%%%%%%%%

\sec{Summary} 

In this article we have calculated transverse energy asymmetries 
as well as optimal observables for \cp\ violating couplings in the 
production and decay of top quarks at the Tevatron upgrade. We have 
compared the physics potential of these observables and 
determined the regions in parameter space with the highest statistical 
significance. It has been found that it is considerable easier to observe \cp\
violation in the $\ttbar$ production process than in
the semi-leptonic decays of the top quarks produced at the Tevatron. 
It has been demonstrated that the single-leptonic modes are are more promising as
signals of \cp\ violation than the dilepton ones.
A more general aim of this paper is 
to point out, that nonstandard \cp\ violation 
in top-quark interactions may be found already before precision 
measurements at the LHC will be done.

\vspace*{0.7cm}
% AAAAAAAAAAAAAAAAAAAAAAAAAAAAAAAAAAAAAAAAAAAAAAAAAAAAAAAAAAAAAAAAAAA
\centerline{ACKNOWLEDGMENTS}

\vspace*{0.3cm}

This
work is supported in part by the Committee for Scientific Research
(Poland) under grant 2\ P03B 180 09, by Maria Sk\l odowska-Curie
Joint Found II (Poland-USA) under grant MEN/NSF-96-252, and by the
Deutsche Forschungsgemeinschaft under grant La 590/2-2. 
We thank Zenro Hioki for his kind help concerning tests of our numerical
analysis. 

\vspace*{0.7cm}
% RRRRRRRRRRRRRRRRRRRRRRRRRRRRRRRRRRRRRRRRRRRRRRRRRRRRRRRRRRRRRRRRRRR


\begin{thebibliography}{99}
\bibitem{DV}
J.F. Donoghue and G. Valencia, \prl{58}{1987}{451}.
%
\bibitem{cprelation} 
W. Bernreuther, O. Nachtmann, P. Overmann, and T. Schr\"{o}der,
\npb{388}{1992}{53};\\
\bg\ and J.F. Gunion, \plb{287}{1992}{237}.
%
\bibitem{cptop}
C.A. Nelson, \prd{41}{1990}{2805};\\
W. Bernreuther and O. Nachtmann, \plb{268}{1991}{424};\\
R. Cruz, \bg\ and J.F. Gunion, \plb{289}{1992}{440};\\
W. Bernreuther and T. Schr\"{o}der, \plb{279}{1992}{389};\\
D. Atwood and A. Soni, \prd{45}{1992}{2405};\\
W. Bernreuther, J.P. Ma, and T. Schr\"{o}der, \plb{297}{1992}{318};\\
A. Brandenburg and J.P. Ma, \plb{298}{1993}{211};\\
\bg\ and W.-Y. Keung, \plb{316}{1993}{137};\\
D. Chang, W.-Y. Keung, and I. Phillips, \prd{48}{1993}{3225};\\
G.A. Ladinsky and C.-P. Yuan, \prd{49}{1994}{4415};\\
B. Ananthanarayan and S.D. Rindani, \prd{51}{1995}{5996};\\
P. Poulose and S.D. Rindani, \plb{349}{1995}{379}, \prd{54}{1996}{4326},
\plb{383}{1996}{212}.
%
\bibitem{CKP}
D. Chang, W.-Y. Keung, and I. Phillips, \npb{408}{1993}{286}.
%
\bibitem{SP}
C.R. Schmidt and M.E. Peskin, \prl{69}{1992}{410}.
%
\bibitem{BG}
\bg, \plb{305}{1993}{384}.
%
\bibitem{AS}
T. Arens and L.M. Sehgal, \prd{50}{1994}{4372};\\ 
B. Lampe, {\it Acta\ Phys.\ Pol.\ }  {\bf B24} (1993), 1079.

%
\bibitem{BGZH}
\bg\ and Z. Hioki, \npb{484}{1997}{17},\\
\plb{391}{1997}{172}.

%
\bibitem{topdisc} F. Abe et al. (CDF Coll.), \prd{50}{1994}{2966}. 


\bibitem{opt-96}
J.F. Gunion, \bg\ and X-G. He, \prl{77}{1996}{5172};\\
see also: \\
D. Atwood and A. Soni, \prd{45}{1992}{2405};\\
M. Davier, L. Duflot, F. Le Diberder and A. Roug\'{e},
\plb{306}{1993}{411};\\
M. Diehl and O. Nachtmann, \zfp{62}{1994}{397};\\
\bg\ and J. Gunion, \plb{350}{1995}{218}.

%
\bibitem{form}
J.A.M. Vermaseren, Symbolic Manipulation with FORM version 2, Tutorial and
Reference Manual, CAN, Amsterdam 1991, ISBN 90-74116-01-9 

%
\bibitem{morfin} 
J.G. Morfin and W.-K. Tung, \zfp{52}{1991}{13}.

%
\bibitem{kleiss} 
S.D. Ellis, R. Kleiss and W.J. Stirling, {\it Comp. Phys. Comm.} {\bf 40}
(1986) 359.  

% 
\bibitem{bernr} 
A. Brandenburg and J.P. Ma, \plb{298}{1993}{211}.

%
\bibitem{upgrade}
R. Frey \etal, ``Top Quark Physics: Future Measurements'', hep-ph/9704243,
April 1997.
\end{thebibliography}
\end{document}